# Why is life so exact?

Semenov D.A. (dasem@mail.ru)

**Abstract.**

This is my attempt to answer Schrödinger's question: "What is Life?". In living cell the local combination of atoms is reproduced with incredible accuracy. In the case of protein biosynthesis the notion of the physical aspect of the process can be qualitatively enriched by taking into account the symmetry of conformational oscillations. This is the key to understand the selectivity in biochemistry.

**Introduction**

In his fundamental book "What is Life? The Physical Aspect of the Living Cell." Erwin Schrödinger asked plainly: "Why does life reproduce itself so exactly?"

His friend and colleague E. Wigner in his study entitled "The probability of the existence of a self-replicating unit" wrote more specifically: "The main idea is to show that in accordance with the quantum-mechanical theory, the probability of the existence of self-replicating units is equal to zero". It is noteworthy that Wigner also stated that the limitations revealed by him can be removed if self-replication requires the symmetry (for example, spin symmetry) to be preserved.

Wigner's study provides a way to interpret Schrödinger's motivation wider: Life does not agree with quantum mechanics, so the study of physical foundations of life must enrich Physics. Moreover, Schrödinger pointed out that life disagrees with statistical physics, which suggests that he saw even greater fundamental significance of the issue. Being an outstanding scientist, Schrödinger, as early as 1943, indicated the area where the scope of science should be broadened.

Having paid a tribute to these remarkable scholars in the statement of the problem, I will hereafter refer to the main question of my work, viz. "What is the physical foundation of molecular biological systems being so incredibly exact and sensitive (selective)?", as Schrödinger-Wigner's problem. In my opinion, my study answers this question and can be useful to physicists, chemists, and biologists.

**Background and the current state of Schrödinger-Wigner's problem**

The magnitude of the effect produced by "What is Life?", written in 1943, cannot be overestimated. The attention of physicists was attracted to biological objects. The boom in biophysics and molecular biology can, in many respects, be accounted for by the social effect of Schrödinger's book. Not only had biology received very much new equipment and numerous experimental opportunities. More importantly, it acquired a new, more developed approach, advanced philosophy.

Yet, until now, more than sixty years after the publication of the book, it has not produced the effect expected by its author. Physics has really enriched biology, but biology has not enriched fundamental physics.

As Schrödinger-Wigner's problem has long been waiting for its solution for biology, I suggest considering my text as an attempt to resolve the fundamental problem of biology and, simultaneously, obtain a solution for physics and chemistry. As will be shown below, the importance of this problem for the three sciences has been significantly underestimated. Current theoretical concepts and experimental potential have reached the necessary level, but remain unused in this area.

**Schrödinger's ideas**

Let me now cite some of the arguments proposed by Schrödinger. This digression is necessary in order to make the statement of the problem that will follow maximally clear. The solution proposed below was anticipated and expected back in 1943. In a sense, it was predetermined. Yet, it may seem unexpected.

The aperiodic crystal concept. Schrödinger summed up the idea of the book in two words – "aperiodic crystal". This is an inherently contradictory combination, indicative of the paradox, internal conflict, the necessity of leaving the boundaries of the existing conceptions. This wording is similar to koans in Zen-Buddhism – short, inherently contradictory phrases encouraging a thinker to alter his attitude to the subject of reflection. Founders of quantum mechanics were the best among physicists at creating physical "koans".

The notion of quasi-crystal and its mathematical prototype – Penrose tiling – have been in use since the 1970s, but by no means can these objects be equated to Schrödinger's "aperiodic crystal". Schrödinger was familiar with the notion of liquid crystals and he realized that periodicity is the essential feature of crystals. The conception of quasi-crystals and Penrose tiling broaden the notion of ordinary crystals, and strict periodicity is replaced by self-similarity, but the main idea – deterministic filling of the whole available space – remains the same. That is, both crystal and quasi-crystal are non-local. Schrödinger's aperiodic crystal is an opposite phenomenon: the local combination of atoms is reproduced with incredible accuracy.

2. *The most essential part of a living cell – the chromosome fibre – may suitably be called an aperiodic crystal. In physics we have dealt hitherto only with periodic crystals.*

For our further discussion we'll need the final chapter (VII) of the book "What is Life?": "Is Life Based on the Laws of Physics?"

62. *What I wish to make clear is that from all we have learnt about the structure of living matter, we must be prepared to find it working in a manner that cannot be reduced to the ordinary laws of physics. And that not on the ground that there is any 'new force' or what not, directing the behaviour of the single atoms within a living organism, but because the construction is different from anything we have yet tested in the physical laboratory.*

64. *The regular course of events, governed by the laws of physics, is never the consequence of one well-ordered configuration of atoms – not unless that configuration of atoms repeats itself a great number of times, either as in the periodic crystal or as in a liquid or in a gas composed of a great number of identical molecules.*

64. *It is not that we can never observe the fate of a single small group of atoms or even of a single atom. We can, occasionally. But whenever we do, we find complete irregularity, co-operating to produce regularity only on the average.*

65. *In biology we are faced with an entirely different situation. A single group of atoms existing only in one copy produces orderly events, marvellously tuned in with each other and with the environment according to most subtle laws.*

*65. Whether we find it astonishing or whether we find it quite plausible that a small but highly organized group of atoms be capable of acting in this manner, the situation is unprecedented, it is unknown anywhere else except in living matter. The physicist and the chemist, investigating inanimate matter, have never witnessed phenomena which they had to interpret in this way.*

*72. That seems very trivial but it does, I think, hit the cardinal point. Clockworks are capable of functioning 'dynamically', because they are built of solids, which are kept in shape by London-Heider forces, strong enough to elude the disorderly tendency of heat motion at ordinary temperature.*

*Now, I think, few words more are needed to disclose the point of resemblance between a clockwork and an organism. It is simply and solely that the latter also hinges upon a solid –the aperiodic crystal forming the hereditary substance, largely withdrawn from the disorder of heat motion.*

### Genetic code translation

The degree of accuracy of protein synthesis (translation) is one mistake per 10000 amino acids. Even in this form the information on the selectivity of translation looks impressive. Biologists are safeguarded against unrelieved surprise at this fact by the replication system operating with the accuracy that is three orders of magnitude higher, with less than one mistake per million copies. Even this does not cause any surprise, as there is a ready explanation that the root of the accuracy is in the complexity of enzyme structure. Thus, the puzzle is explained through the mystery and stops exciting the minds of the majority. This is a natural, maybe even useful, response to an unsolvable problem – one tries not to notice it, not to upset oneself.

None of the ever developed chemical (non-enzymatic) ways to synthesize biopolymers, either proteins or nucleic acids, is sufficiently selective. Such high selectivity is unknown even for organic reactions in solution.

The requirement of an arbitrarily high closeness to 100% accuracy could be somewhat slackened if there were ways to correct the mistakes. Fortunately, no such processes are known for translation, which is its advantage over replication. "Fortunately" and "advantage" certainly mean that this enhances the clearness of this study.

In his study [2] Wigner united (mixed) three essentially different issues: the accuracy of operations in molecular systems, the possibility of self-reproduction, and the possibility of spontaneous generation of life. Wigner was impressed by von Neumann's work [3] and proposed arguments against his idea of the possibility of self-reproduction.

Protein synthesis is not directly related to the issues of self-reproduction and generation of life. Hence, I will further dwell on just one aspect of Schrödinger-Wigner's problem. However, after considering the mechanism of protein synthesis, I am going to show that the obtained solution is significant in a much more general case.

In my previous studies [4, 5] I proved that, firstly, it is the form, or, more exactly, conformation, of the first two nucleotides of the codon that is recognized. Secondly, the impact significant for recognition can be smaller than the energy of one hydrogen bond.

These statements are both directly related to Schrödinger-Wigner's problem. The first is important because fluctuations of the polymer chain must disorder (destabilize) conformation in the solution. It cannot be considered as something stable. This thermodynamic limitation seems quite rational. The second aspect – sensitivity to interactions that are weaker than the energy of a hydrogen bond – strengthens the first statement. Moreover, it becomes clear that the energy of this impact is the effect at the quantum level: the chemical bond is behavior of an electron, the electron is a quantum object, and quantum mechanics cannot provide such accuracy.

**Von Neumann's approach**

In his study [6], von Neumann proposed a possible way to enhance the reliability of the operation of automatic devices consisting of unreliable components. In the simplest case, reliability can be enhanced by dual redundancy of components and units, while in a more general case – by making the system more complex. As von Neumann studied information processing, I will not discuss his work in detail here.

A similar approach is proposed by the information theory [7], but the emphasis is placed on the message rather than on the automatic device. In the simplest case, the reliability of the message is enhanced by repeating the message. In a more complicated variant, the content is spread (distributed) along the message, the structure of the message becomes more complicated, but this turns out to be a more economic approach. Please note that the second approach necessarily involves decoding, as a way to recover the original content based on the impaired signal.

Biologists implicitly assume that the reason for selectivity is concealed in the intricate structure of the decoding system, i.e. in the complex translation system. This is the way biology has interpreted von Neumann's and Shannon's metaphors. But where is the physical mechanism of decoding and what is its essence?

**Physical mechanism**

What does the mechanism of recognition consist in? If recognition is just a complementary interaction, in what way is the system further directed towards a new state? The answer sanctified by thermodynamics is that the system minimizes the energy, i.e. accepts the most advantageous conformation.

Please note that none of the components of the translation system, except aminoacyl tRNA, are codon-dependent, and, hence, they cannot be directly responsible for the selectivity of the process. Thus, we have nothing but tRNA.

In my previous studies [4, 5], I made an assumption that sensitivity of the translation system is based on a specific physical state, similar to staying near the critical point of the second-order phase transition. The physical theory suggests that in this case diffusion stops, as if the thermodynamic system is transformed into the classical one. Note that molecular movement does not stop; the movement of the particles gets more and more interrelated and coordinated. Interactions ignored under ordinary conditions become significant; the leading part is played by the symmetry of the system rather than by energy.

It is noteworthy that phase transitions are used in chemistry to enhance selectivity of certain processes.

This indication of the necessity for the search for symmetry can be specified further: a) there had been no symmetry before interaction of tRNA with the codon; b) it resulted from this interaction; c) the subsequent symmetry reduction results in a new state. At the same time, the system must not be a quantum one, i.e. this is neither the Jahn-Teller effect nor the Woodward-Hoffmann rules. One can easily see that the search for locally symmetric sites in the structure of the tRNA-codon system is useless. The "non-quantumness" is provided by structures bulkier than electrons – atoms, monomer components, and oligomeric segments – being involved in the arrangement of the symmetry.

Note that in the spatial structure of tRNA there are two sites, the Anticodon arm and the T arm, which are very similar to each other in the shape of the sugar-phosphate backbone.

a) Prior to interaction with the codon, the helix of the anticodon loop is not bound to anything, and the T arm interacts with the D arm to form complementary pairs of two nucleotides.

b) As a result of interaction with the codon, both sites get fixed in nearly identical positions. The sites become identical in that their conformational oscillations can now be synchronized. Synchronous oscillations can be regarded as symmetry of the system.

c) The symmetry reduces spontaneously as the oscillation energy reaches a certain level, suggesting that oscillations are an accumulator of energy. The energy that has accumulated in oscillations is released, causing large-scale changes in tRNA conformation. What makes this similar to the second-order phase transition is the decisive role of symmetry and the maximal loss of redundant degrees of freedom at the point of phase transition.

Thus, point a) and Figure 1 show that there is no symmetry in the system. In point b), two similar sites arise: two helixes formed by the pins of the anticodon loop (purple) and the Tψ loop (pink), both fixed by the complementary interaction organized by three and two nucleotides, respectively.

Conformational movements, based on rotational degrees of freedom, must cause these helixes to twist. The necessary symmetry is not a static feature; symmetry suggests the possibility of synchronous oscillations. Synchronism implies equivalence, in a sense.

It is noteworthy that the two helix segments are similar but not strictly identical, suggesting that not every oscillation mode can be synchronized. Limitations placed by the properties of one helix on oscillations of the other lead to selection and mutual filtration of oscillations, i.e. to a decrease in the possible number of the degrees of freedom.

One of the results of thermodynamic examination of polymeric chains is the finding that linear polymers cannot but be subjected to significant fluctuations. That is, a fluctuation accumulating considerable energy must occur in the system within a relatively short time. Most probably, this fluctuation must be transformed into synchronous oscillations, i.e. it will be subjected to filtration, standardization, and limitation of the degrees of freedom.

If the energy of oscillations is too large, the system can collapse. In the case with the tRNA, this will mean the breaking of the hydrogen bonds between the site of the T arm and the site of the D arm. The energy accumulated in synchronous oscillations must be released, causing large-scale changes in tRNA conformation.

It does not always make sense to wait for a great fluctuation. If conformational changes require considerable energy expenditures, the energy can be recovered from chemical reactions. The most significant reaction of this kind is hydrolysis of phosphodiester bond. Similarly to the energy of a large fluctuation, the energy of chemical reaction is transformed into the movements of the parts of the molecules, causing generation of synchronous oscillations, a reduction in the number of the degrees of freedom, and a subsequent reduction in the symmetry of the system. It is important that mere expenditure of ATP cannot provide selectivity of the process. If the energy is spent on the amplification of random intramolecular oscillations, the selectivity of the process can only decrease.

The similarity to the second-order phase transition consists in the necessity of reducing the symmetry, but that is not all. The expansion of the correlation radius, the necessity of taking into account large fluctuations, and the cessation of diffusion also have their analogs in the molecular system considered in this study. In fact, as the transition point gets nearer, intramolecular movements gradually get more correlated, redundant degrees of freedom are suppressed, and the static system is gradually transformed into a dynamic one.

The conception proposed here has an important advantage for experimenters – to confirm it, one "merely" has to record the predicted synchronous intramolecular conformational oscillations.

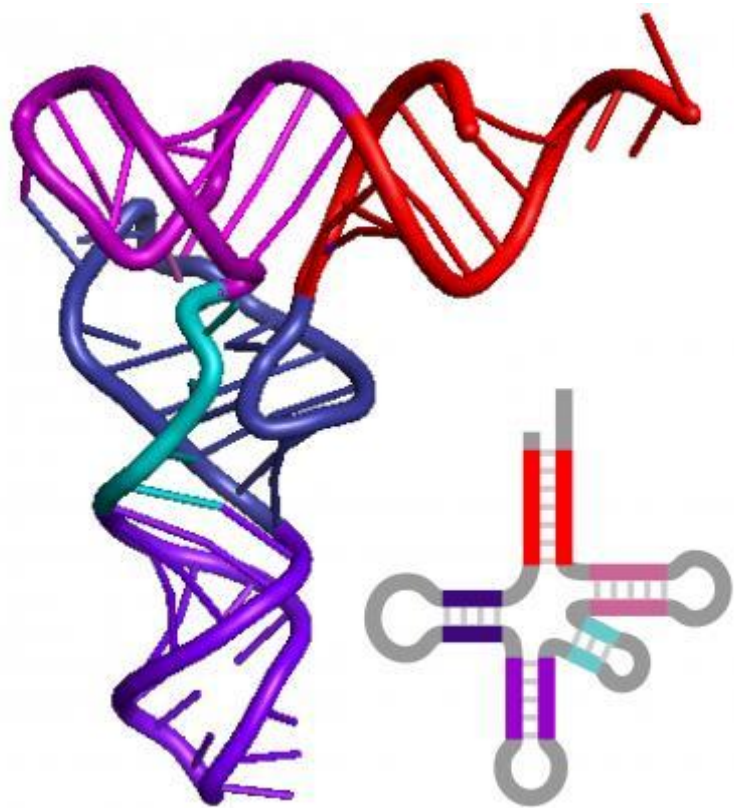

**Figure 1. Structure of tRNA**

**A useful association.**

I cannot help making a brief digression, caused by Schrödinger's association between the dynamic systems and pendulum clocks. The mechanism that I propose is visually more similar to a ribbon in the hands of a rhythmic gymnast than to a pendulum. The ribbon at rest has a great number of degrees of freedom and looks just uncontrollable. The wave motion reduces the number of degrees of freedom and makes the system relatively simple (for skilled gymnasts and experienced referees, but not for us, mere mortals). The energy that is imparted to this system when oscillations are created is sufficient for the system to lose a considerable number of its degrees of freedom. The difference between molecular machines and this ribbon is that there is no gymnast. Most of the audience would be disappointed by the absence of the gymnast, but I hope that the representatives of natural sciences would enjoy the peculiar beauty of this spectacle.

**The role of nonlinearity and discreteness**

For the possible subsequent examination of the predicted process, it is significant that the considered system is nonlinear and discrete. The nonlinearity suggests the general result – the origination of one oscillation mode that consumes the energy of the other modes and subordinates all redundant degrees of freedom. Laser is a well-known illustration. The general principle, however, can be understood from a perfectly nonphysical idea: low intensities can be sufficiently well described linearly, but as the intensity increases, the description has to involve more and more nonlinearities, for example, more and more members of the Taylor series.

The discreteness of the system cannot be ignored either. As demonstrated in the classical Fermi-Pasta-Ulam experiment [8], arrangement of discrete elements in a chain almost generally leads to nonlinear description and predominance of a small number of degrees of freedom. Modern physics and mathematics have made good progress in theoretical understanding of this event. Accurate analytical results have been obtained. I can refer here to studies performed by Toda [9] and by Novikov and Krichever [10]. Although the current state of research in this field does not allow extension of results to the new, entirely non-abstract, object – biological molecules, it is important that the necessary tool is ready! Some of the results obtained in this field are quite general, such as the non-uniform distribution of energy among degrees of freedom in chains.

I should specially note that the Toda chains demonstrate an even closer relationship to the notion of symmetry than I have shown in my text. This fact suggests that dynamic symmetry of tRNA oscillations could be a simplest case of this event in biology.

Please note that the role of nonlinearity in the task of self-replication was mentioned by von Neumann. Even in Schrödinger's book, although he compares living organisms to crystals based on the criterion of the accuracy of reproduction, accuracy in crystals is an essentially nonlinear phenomenon.

**Oscillation as multiple repetitions of the message**

There are no multiple repetitions in the mRNA sequence. There are no indications of the translation system using information from the mRNA segment that is larger than the codon to be read. That is, it would seem that there is no way to enhance the accuracy of message reading, as suggested by von Neumann's and Shannon's ideas.

However, every time step of oscillations can be regarded as reception of the same message, and although the energy of oscillations grows, the message must be read in the same way (i.e. the codon-tRNA complex must not be broken). This multiple reading alone may be considered as a means of enhancing the accuracy.

**Molecular recognition**

The issue of selectivity of biochemical processes is in many respects synonymous to the issue of molecular recognition of substrate. It is traditionally reduced to assumptions regarding the structural, spatial arrangement of the substrate binding region. In Fischer's lock-and-key concept (1913) [11], substrate is bound in the "pocket" of the corresponding shape. This is the first appearance of the notion of complementarity. Koshland's "glove" model (1958) [12] takes into account the necessity of preventing the functioning when similar but smaller substrates are bound. The views of the physical nature of enzyme functioning evolved from the function of substrate passive convergence to the necessity to take into account conformational changes in the course of reaction. In these very productive and certainly important studies, the main stages of the process are considered in terms of thermodynamics, so Schrödinger-Wigner's problem is not resolved. I'll try to discuss differences between such molecular recognition and current concepts.

The conception of intramolecular oscillations has significant advantages over the currently existing notions of molecular recognition. Contemporary views of molecular recognition require noncovalent bonding of the recognized molecule and subsequent transition of the whole complex into a more advantageous, in terms of energy, state. That is, the recognizing molecule has a certain amount of surplus energy, which it cannot expend, prior to substrate binding. Molecular recognition is considered to be, in its essence, a transition to a more stable state with lower energy.

The most significant distinction of my conception is dynamic instability of the symmetric state. In the proposed conception of intramolecular synchronous oscillations, the energy of the system gradually increases after the binding, due to activation of oscillations, or it does not increase if the substrate has not been chosen correctly, or, in the case of the wrong substrate, as the oscillations are activated, the energy is expended on the destruction of the bond with the substrate. This system of substrate recognition has qualitatively new features: for example, it is incapable of accidental activation, because it does not have the necessary energy store. One can say that molecular recognition turns into an active process: recognition and functioning are not assigned to different stages, but constitute a single entity. So, the search for the structures "controlling the correctness of recognition" becomes meaningless. The "external" control of this kind is just contradictory. In terms of quantum mechanics, it is equivalent to measurement of values without affecting the quantum system. In terms of thermodynamics it is analogous to Maxwell's demon. In the proposed conception, the very process of the functioning of the biomolecule is the controlling of the correctness of recognition.

Von Neumann's idea of enhancing reliability by making the construction of the device more complex can also be developed somewhat. The message physically inserted into the molecular mechanism of reading can lead to its qualitative restructuring only if it is true, or, more exactly, acceptable. The criterion of acceptability can be defined very precisely as the large molecular construction can be altered in various minor respects, for example, tRNA nucleotides can be modified.

**Conclusions**

To sum up: The accuracy of the process of protein biosynthesis and other related processes of biosynthesis cannot be accounted for by traditional concepts.

In the case of protein biosynthesis (translation) the notion of the physical aspect of the process can be qualitatively enriched by taking into account the symmetry of conformational oscillations.

This symmetry is dynamical, and the contemporary experimental and theoretical approaches to biological molecules are not adequately prepared for the work with this object.

My study predicts the occurrence of certain oscillations during the functioning of the translation system; so, my hypothesis can be verified. The hypothesis allows experimental verification.

In the second half of the 20th century, mathematics and theoretical physics made a number of systematic steps towards the insight into such dynamic systems. The energy of oscillations in such a system is non-uniformly distributed among degrees of freedom; moreover, the increase in the energy level causes a decrease in the number of degrees of freedom.

Diffuse processes are suppressed, the system functions as a classical one, in spite of its small size. In the limit, the system becomes a deterministic, completely correlated one, which makes its description similar to the description of behavior near the critical point of the second-order phase transition. Another significant point in the description of the system in the given state is a reduction in symmetry rather than energy decrease.

Oscillations are similar to repetitions of the message, enhancing the accuracy of reading.

The introduction of dynamic notions into the conception of molecular recognition enriches it qualitatively. Molecular recognition turns into an active process, capable of precise adjustment and insensitive to random disturbances such as molecules similar to the substrate. This behavior can be termed as "resonance molecular recognition".

**The significance of this study in the fields other than biosynthesis**

It is only natural to think that the behavior described in the case study of tRNA will occur in other molecular devices. Active involvement of oscillations may be expected in artificial ribosomes. Therefore, I have to note that the search for such symmetry in proteins can be very difficult because of the diversity of amino acids and different sizes of the enzyme. However, I will not dwell on this in greater detail as there are several cases whose relationship to Schrödinger-Wigner's problem is not evident but, yet, very important. These are the cases that I'd like to discuss at the end of this narration.

*The Leventhal paradox.* The deterministic structure of proteins is in contradiction to models of statistic physics of polymers, which are successfully used outside biology. The methods currently used to model the arrangement of the protein globule ignore dynamic effects, reducing everything to energy minimization. The occurrence of the patterns that are capable of synchronous oscillations followed by restructuring and symmetry reduction can be a reason for the origin of strict intramolecular order. Moreover, in the course of biosynthesis such structures can occur repeatedly.

*The action of chaperones.* This example naturally follows from the above paragraph and the discussion of molecular recognition. Correct shaping of proteins is possible not only because there are symmetric oscillations inside the molecule but also due to the occurrence of oscillations after the molecule is incorporated into a symmetric supramolecular system, capable of synchronous conformational oscillations. This can provide insight into the mechanism of action of chaperones.

*Molecular recognition in chemistry.* The insight into the nature of biological "molecular recognition" can enrich chemistry considerably. The prospect of overcoming the fundamental limitations of thermodynamics and quantum mechanics is certainly very tempting. If we restrict ourselves to "molecular recognition", i.e. the binding to a certain substrate and the obtaining of "response" to it, we can be more exact. Nature often makes its solutions bulkier and less evident than a scientist would. For the symmetric situation to occur after binding with the substrate, it is natural to propose incorporating one molecule of the substrate (a template) into the recognizing system. This effect may also underlie enzyme activation by the substrate.

*Microelectronics.* As the components of processors get smaller, Schrödinger-Wigner's problem becomes increasingly important for microelectronics. Limitations imposed by quantum mechanics and thermodynamics must start interfering with the operation fail-safety. Physics and technology have resolved very many issues better than nature has: the efficiency of photosynthesis is lower than that of solar cells; movement on wheels is more efficient than walking. Yet, physics cannot boast of large-scale precise reproduction of aperiodic structures with 1-nm components. Nature outclasses physics and technology by two orders of magnitude, measured linearly! This means that by understanding the tricks of life microelectronics can succeed in reducing this lag or, at least, determine the possible limits of its own development.

**Acknowledgements.**

I am sincerely grateful to Professor S.P. Gabuda, from whose published studies and personal communication I first learned about a possible relationship between second-order phase transitions and fundamental biological phenomena.

I'd like to express my gratitude to Professor V.O. Bytev, who acquainted me with Toda's works and related studies.

E.V. Chernikov, a friend of mine, was the first to point out the fact that oscillations can be interpreted as repetitions of the message, and I'd like to express my gratitude to him.

I would like to thank Krasova E. for her assistance in preparing this manuscript.

**References:**
1. Schrödinger E. What is Life? The Physical Aspect of the Living Cell. Cambridge at the University Press. 1944.
2. Wigner. E.P. Symmetries and Reflections. Scientific Essays. Indiana University Press, Bloomington, 1967. ch.11
3. von Neumann J., Burks A.W. Theory of Self-Reproducing Automata. Univ. of Illinois Press. 1966.
4. Semenov D.A. Wobbling of What? arXiv:0808.1780
5. Semenov D.A. Evolution of the genetic code. From the CG- to the CGUA-alphabet, from RNA double helix to DNA. arXiv:0805.0484
6. von Neumann J. Probabilistic Logic and Synthesis of Reliable Organisms from Unreliable Components, in Automata Studies, Ed. by C. E. Shannon and J. McCarthy Princeton Univ. Press 1956
7. Shannon C.E., Weaver W. The Mathematical Theory of Communication. Univ of Illinois Press, 1949.
8. Fermi E., Pasta J., and Ulam S., Studies of Nonlinear Problems. American Mathematical Monthly. Vol. 74. №1. 1967.
9. Selected Papers of Morikazu Toda (Series in Pure Mathematics). Edited by Miki Wadati (University of Tokyo). World Scientific Oct. 1993.
10. Krichever I. M., Novikov S.P. Two-dimensionalized Toda lattice, commuting difference operators, and holomorphic bundles, UMN, 2003, 58:3(351), 51–88
11. Fischer E, «Einfluss der Configuration auf die Wirkung der Enzyme» Ber. Dt. Chem. Ges. 1894 v27, 2985—2993.
12. Koshland D.E., Application of a Theory of Enzyme Specificity to Protein Synthesis. Proc. Natl. Acad. Sci. U.S.A. 1958 Feb;44(2):98-104.